\documentclass[preprint,5p,times,twocolumn]{elsarticle}

\usepackage{amsmath}
\usepackage{amssymb}
\usepackage{mathtools}
\usepackage{graphicx}
\usepackage{subcaption}
\usepackage{booktabs}
\usepackage{siunitx}
\usepackage{xcolor}
\usepackage{tabularx}
\usepackage{lineno}
\usepackage[hidelinks]{hyperref}


\graphicspath{{./}}

\journal{}

\begin{document}

\begin{frontmatter}

\title{Machine-learning-assisted multiscale topology optimization of functionally graded superimposed lattice structures}

% Author information
\author[inst1]{Prashant Kumar Gupta}
\ead{prashant_kg@ce.iitr.ac.in}

\author[inst2]{Jonathan Stollberg}
\ead{jonathan.stollberg@tu-darmstadt.de}

\author[inst2]{Dominik Schillinger}
\ead{dominik.schillinger@tu-darmstadt.de}

\author[inst1]{Mohammad Ashraf Iqbal}
\ead{ashraf.iqbal@ce.iitr.ac.in}

\affiliation[inst1]{
    organization={Department of Civil Engineering, Indian Institute of Technology Roorkee},
    city={Roorkee},
    postcode={247667},
    state={Uttarakhand},
    country={India}
}

\affiliation[inst2]{
    organization={Institute for Mechanics, Computational Mechanics Group, Technical University of Darmstadt},
    city={Darmstadt},
    postcode={64287},
    country={Germany}
}

\begin{abstract}
Functionally graded lattice structures enable lightweight designs with spatially tunable stiffness and density, but their use in multiscale topology optimization is limited by the cost of repeated computational homogenization. This work presents a machine-learning-assisted multiscale optimization framework for regular superimposed lattice structures. The unit cell is formed by combining body-centered cubic, face-centered cubic, and simple cubic lattice components, each controlled by an independent geometric parameter. Offline computational homogenization is used to generate effective stiffness data, which are then used to train a Cholesky-constrained neural network surrogate. This representation reconstructs the homogenized stiffness tensor in a physically admissible form. A separate neural network is trained to predict relative density from Monte Carlo-based density estimates. We incorporate our surrogates into a two-stage topology optimization strategy. First, a macroscale topology is obtained using the solid isotropic material with penalization (SIMP) method. The resulting solid region is then used for microscale lattice optimization, where the local lattice parameters are updated using the method of moving asymptotes (MMA). The trained stiffness and density surrogates replace repeated online homogenization during this stage. The method is demonstrated on a three-dimensional Messerschmitt-B\"olkow-Blohm (MBB) beam benchmark, producing spatially varying lattice parameters and relative density fields consistent with compliance minimization under a material constraint.
\end{abstract}

\begin{keyword}
Multiscale topology optimization \sep Functionally graded lattice structures \sep Physics-augmented neural networks \sep Additive manufacturing \sep Computational homogenization
\end{keyword}

\end{frontmatter}

% ============================================================
\section{Introduction}
\label{sec:introduction}
% ============================================================

\begin{figure*}[t]
    \centering
    \includegraphics[width=0.95\textwidth]{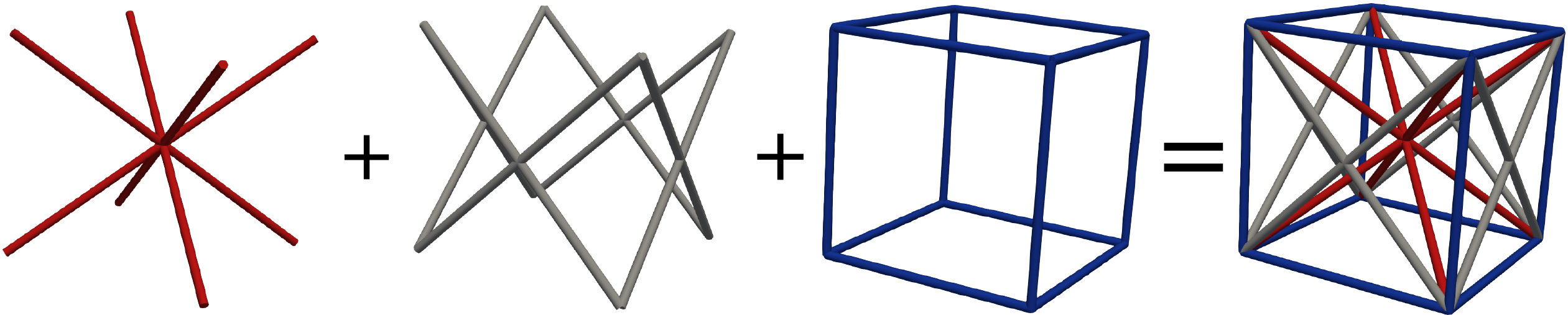}
    \caption{Construction of the regular superimposed lattice unit cell by combining BCC (red), FCC (gray), and simple cubic (blue) lattice components. The parameters $a_1$, $a_2$, and $a_3$ control the corresponding BCC, FCC, and simple cubic contributions, respectively.}
    \label{fig:superimposed_lattice}
\end{figure*}

Topology optimization provides a systematic framework for distributing material within a prescribed design domain to improve structural performance. Among density-based methods, the solid isotropic material with penalization (SIMP) approach remains one of the most widely used formulations because of its simplicity, compatibility with finite element analysis, and ability to generate clear load-carrying layouts \cite{Bendsoe_2004,Sigmund_2001,Sigmund_2013}. However, conventional SIMP-based optimization is essentially a single-scale method: it determines where material should be placed, but it does not directly control the microstructural architecture of the material.

This limitation is important for lattice-based structures, where the effective stiffness, density, anisotropy, and manufacturability depend strongly on the unit-cell geometry. Lattice cells such as body-centered cubic (BCC), face-centered cubic (FCC), cubic, octet, and other architected cells have been widely studied for lightweight structural design, in particular in combination with additive manufacturing \cite{Gibson_2015,Zhu_2021}, because their mechanical response can be tuned through topology and geometric parameters \cite{Deshpande_2001,Maskery_2018,Kolken_2017,Chibinyani_2024}. Functionally graded lattice structures extend this idea by allowing unit-cell parameters to vary spatially, so that stiffer or denser microstructures can be assigned to critical load-carrying regions while lighter configurations are used elsewhere \cite{Panesar_2018,Niknam_2020}.

Multiscale topology optimization, pioneered by the homogenization method of Bends\o{}e and Kikuchi \cite{Bendsoe_1988}, provides a natural route for coupling structural-scale material layout with microscale material architecture. Both concurrent and sequential multiscale formulations have been studied extensively \cite{Wu_2021,Allaire_2002,Coelho_2008,Groen_2018,Gangwar_2021}. A computationally attractive class of methods performs homogenization-based topology optimization over lattice unit cells of predefined geometry that are parametric only in their relative density or a few geometric parameters, which keeps the number of design variables small and remains an active field of research \cite{Wang_2017,Li_2018,Stromberg_2024}. In these methods, homogenization is commonly used to link the microscale representative volume element (RVE) to the macroscale continuum response \cite{Guedes_1990,Zaoui_2002,Zohdi_2004,Andreassen_2014}. Although homogenization avoids direct simulation of the full lattice structure, it becomes computationally expensive when embedded inside an optimization loop, since the local lattice parameters may change repeatedly over many elements and design iterations.

Surrogate modeling offers a practical way to reduce this computational cost. Neural networks have increasingly been used to approximate the mapping from microstructural parameters to effective material properties in multiscale and lattice-based topology optimization \cite{Lu_2020,Li_2022,Yu_2023,Gartner_2021,Woldseth_2022,Wang_2022,Bai_2024,Bessa_2017}. Once trained, these models can provide fast stiffness and density predictions and can be embedded into gradient-based optimization frameworks. However, directly predicting stiffness matrix entries can produce non-physical tensors if symmetry and positive definiteness are not enforced. Since the predicted stiffness tensor enters the finite element equilibrium equation, physical admissibility is important for stable and meaningful structural simulations.

Physically constrained neural network material models address this by enforcing physical requirements directly in the model architecture rather than through the training loss alone. Representative families include physics-informed \cite{Raissi_2019}, physics-augmented \cite{Klein_2022,Linden_2023}, and constitutive \cite{Linka_2021} neural networks. In particular, Cholesky-based or physics-augmented parameterizations reconstruct the stiffness tensor from constrained factors rather than predicting it entry-by-entry \cite{Shojaee_2024,Stollberg_2025, Stollberg_2026}, which helps preserve symmetry and positive definiteness of the predicted effective stiffness. Related multiscale lattice optimization frameworks have also shown the value of combining offline homogenization data with neural network surrogates for efficient structural optimization \cite{Stollberg_2025}.

In this work, we develop a machine-learning-assisted multiscale topology optimization framework for functionally graded superimposed lattice structures. The microscale material is a regular unit cell formed by superimposing BCC, FCC, and simple cubic lattice components, whose three independent geometric parameters provide an interpretable and continuous design space for locally tuning stiffness and density. To avoid repeated online homogenization inside the optimization loop, the effective stiffness and relative density of this lattice family are represented by two trained neural network surrogates, which are embedded in a two-stage macroscale--microscale optimization strategy.

The main contributions of this work are:
\begin{enumerate}
    \item A regular superimposed lattice unit cell is introduced by combining BCC, FCC, and simple cubic lattice components.
    \item A Cholesky-constrained neural network surrogate is used to predict homogenized stiffness tensors in a physically admissible form.
    \item A separate density surrogate is trained using Monte Carlo-based relative density estimates, which are suitable for superimposed lattices with overlapping struts.
    \item The stiffness and density surrogates are integrated into a two-stage multiscale topology optimization framework.
    \item The framework is demonstrated on a three-dimensional benchmark problem, producing spatially varying lattice parameters and relative density fields.
\end{enumerate}

The rest of the paper is organized as follows. Section~\ref{sec:surrogates} presents the machine-learning part of the framework, covering the superimposed lattice unit cell, the homogenization procedure and dataset, and the stiffness and density surrogate models with their training and validation. Section~\ref{sec:optimization} presents the two-stage multiscale topology optimization formulation. Section~\ref{sec:results} reports the computational results, covering the macroscale topology optimization, the microscale lattice optimization and its convergence, and the resulting optimized lattice parameter and density fields, followed by a discussion. Section~\ref{sec:conclusions} summarizes the main findings and outlines future extensions.

% ============================================================
\section{Machine-learning surrogate modeling of the superimposed lattice material}
\label{sec:surrogates}
% ============================================================

The microscale material of the proposed framework is represented by a regular superimposed lattice unit cell. Its effective stiffness and relative density are predicted using trained neural network surrogate models, so that repeated computational homogenization is not required during the optimization loop.

The framework consists of four main parts: (i) definition of the superimposed lattice unit cell, (ii) computational homogenization of the lattice unit cells, (iii) training of stiffness and relative density surrogate models, and (iv) two-stage topology optimization. This section describes the modeling and machine-learning parts (i)--(iii), each together with its numerical implementation and corresponding results.

% ------------------------------------------------------------
\subsection{Superimposed lattice unit cell}
\label{subsec:lattice_cell}
% ------------------------------------------------------------

\begin{figure*}[t]
    \centering
    \includegraphics[width=0.7\textwidth]{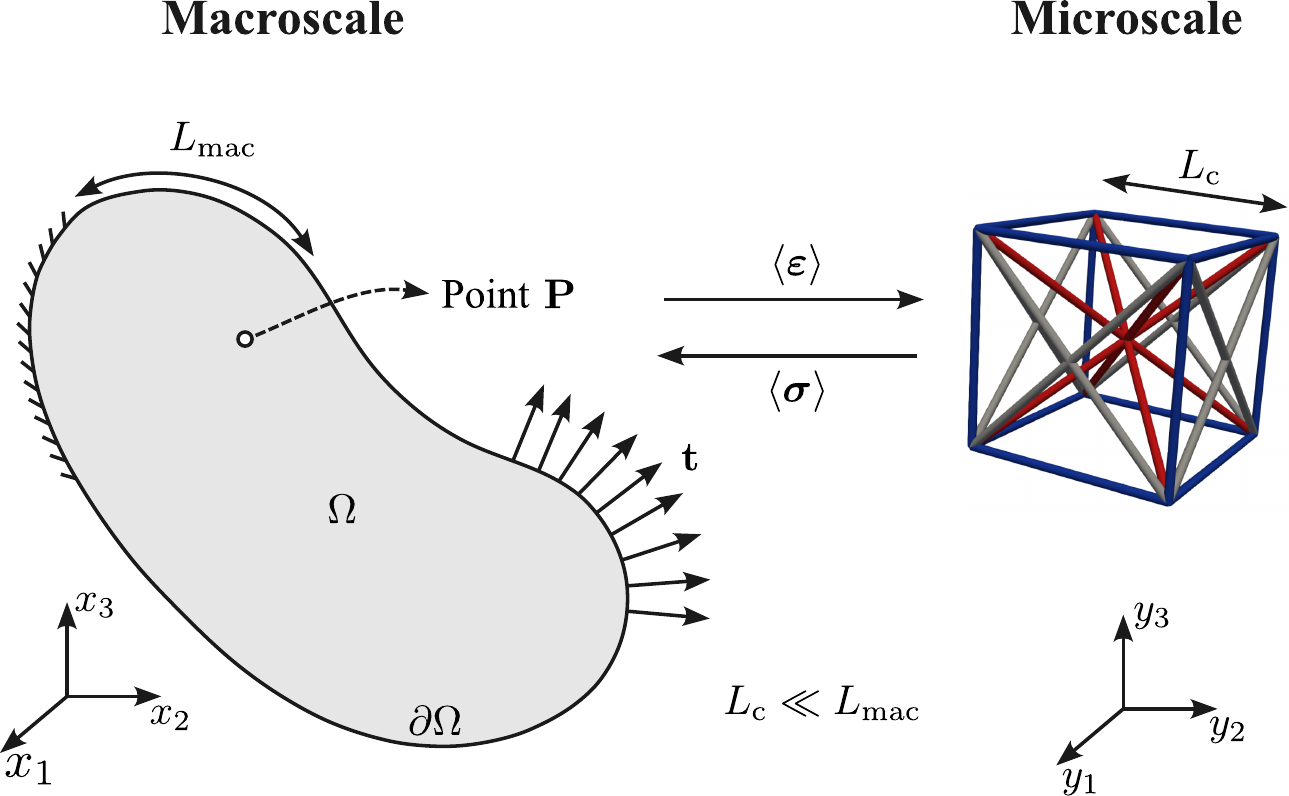}
    \caption{Multiscale homogenization concept linking the macroscale continuum problem to the microscale superimposed lattice unit cell \cite{Gangwar_2021}. The macroscale body occupies the domain $\Omega$ with boundary $\partial\Omega$, on which a displacement or a surface traction $\mathbf{t}$ may be prescribed. At a material point $\mathbf{P}\in\Omega$, the macroscopic strain $\langle\boldsymbol{\varepsilon}\rangle$ is imposed on the unit cell, and the macroscopic stress $\langle\boldsymbol{\sigma}\rangle$ is recovered as the volume average of the microscale stress over the cell. The vectors $\mathbf{x}=[x_1,x_2,x_3]^{\mathrm{T}}$ and $\mathbf{y}=[y_1,y_2,y_3]^{\mathrm{T}}$ denote the macroscale and microscale (unit-cell) coordinates, respectively.}
    \label{fig:multiscale_homogenization}
\end{figure*}

The representative unit cell used in this work is constructed by superimposing three regular lattice topologies: body-centered cubic (BCC), face-centered cubic (FCC), and simple cubic, as illustrated in Fig.~\ref{fig:superimposed_lattice}. The three components are controlled independently through the design variable vector
\begin{align}
    \mathbf{a} = [a_1, a_2, a_3]^{\mathrm{T}} \,,
    \label{eq:lattice_parameters}
\end{align}
where the parameters $a_1$, $a_2$, and $a_3$ correspond to the BCC, FCC, and simple cubic components, respectively. Each $a_i$ is defined as the aspect ratio of the strut diameter $d_i$ of the corresponding lattice component to the unit-cell edge length $L_\mathrm{c}$,
\begin{align}
    a_i = \frac{d_i}{L_\mathrm{c}} \,,\qquad i=1,2,3 \,,
    \label{eq:aspect_ratio_definition}
\end{align}
so that $a_i$ is a dimensionless measure of the actual material infill of the unit cell. We emphasize that the aspect ratio is closely related to the relative density $\tilde{\rho} = \tilde{\rho} ( \mathbf{a} )$ of the unit cell, that is the ratio of solid (strut) volume to the total unit cell volume $L_{\mathrm{c}}^3$.

The purpose of using a superimposed lattice is to obtain a continuous and interpretable microstructural design space. Instead of selecting a single unit cell type at each material point, the local stiffness and density can be tuned by changing the relative contribution of the three lattice components. This is useful for functionally graded lattice structures because different regions of the same structure may require different effective material responses.

% ------------------------------------------------------------
\subsection{Computational homogenization of lattice unit cells}
\label{subsec:homogenization}
% ------------------------------------------------------------

\begin{figure*}[t]
    \centering
    \includegraphics[width=0.95\textwidth]{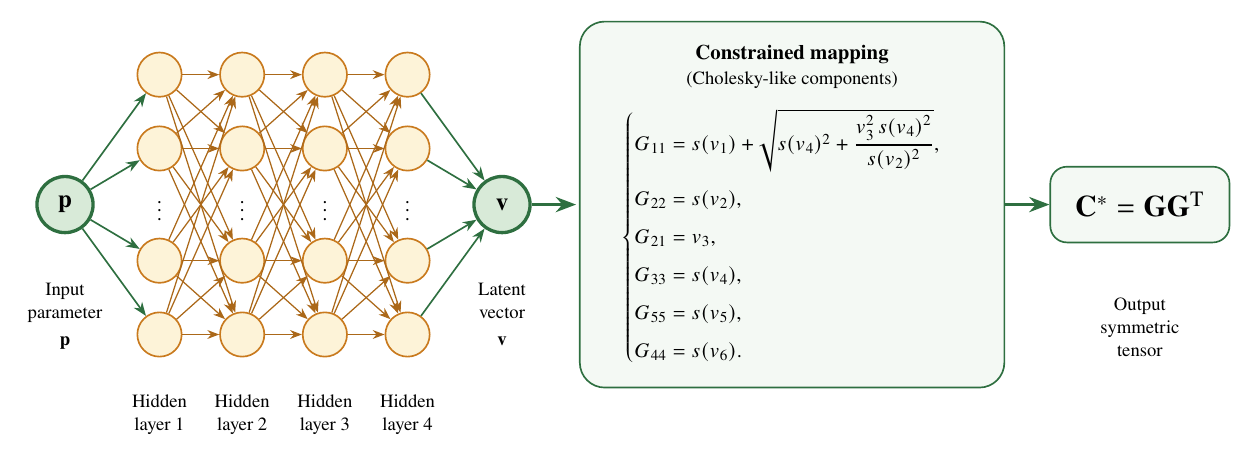}
    \caption{Cholesky-constrained stiffness surrogate used to predict the homogenized stiffness tensor. The neural network maps the input vector $\mathbf{p}=[a_1,a_2,a_3,E,\nu]^{\mathrm{T}}$ to a latent vector $\mathbf{v}$, which is transformed into physically admissible Cholesky-like components. The effective stiffness tensor is reconstructed as $\mathbf{C}^{*}=\mathbf{G}\mathbf{G}^{\mathrm{T}}$.}
    \label{fig:cholesky_nn}
\end{figure*}

The effective material response of each lattice configuration is obtained using computational homogenization \cite{Zohdi_2004}. Throughout this work, the analysis is restricted to linear (small-strain) elasticity. Because the lattice is periodic, the RVE reduces to a single periodic unit cell, whose domain we denote by $\omega$ and which serves as the microscale domain throughout this work. For a given set of lattice parameters $\mathbf{a}$ and an isotropic strut base material with Young's modulus $E$ and Poisson's ratio $\nu$, a unit cell is generated so that its homogenized response is fully characterized by an effective linear-elastic stiffness tensor.

The homogenization procedure follows the usual scale-separation assumption,
\begin{align}
    L_{c} \ll L_{\mathrm{mac}} \,,
    \label{eq:scale_separation}
\end{align}
where $L_\mathrm{c}$ is the unit-cell edge length introduced in Eq.~\eqref{eq:aspect_ratio_definition} and $L_{\mathrm{mac}}$ is the characteristic length of the macroscale structure. Under this assumption, the microscale lattice can be replaced by an equivalent continuum material at the macroscale, as illustrated in Fig.~\ref{fig:multiscale_homogenization}.

The superimposed lattice unit cells are generated using an in-house homogenization implementation from a related multiscale lattice optimization framework \cite{Stollberg_2025}. For each sampled parameter vector
\begin{align}
    \mathbf{p} = [a_1, a_2, a_3, E, \nu]^{\mathrm{T}} \,,
    \label{eq:stiffness_input}
\end{align}
the struts of the BCC, FCC, and simple cubic components are created and meshed as one-dimensional entities in a three-dimensional physical space using Gmsh \cite{Geuzaine_2009}. The resulting mesh is converted into an Abaqus input file where we assign cubic Euler-Bernoulli beam elements to the discretization. Furthermore, struts are connected through rigid joints. We then apply periodic boundary conditions on opposite faces of the unit cell,
\begin{align}
    \mathbf{u}^{+} - \mathbf{u}^{-} = \langle \boldsymbol{\varepsilon} \rangle \, (\mathbf{y}^{+}-\mathbf{y}^{-}) \,,\qquad \boldsymbol{\theta}^{+} - \boldsymbol{\theta}^{-} = \mathbf{0} \,,
    \label{eq:pbc}
\end{align}
where $\mathbf{u}^{\pm}$ are the displacements on opposing boundary faces, $\mathbf{y}^{\pm}$ are the corresponding nodal positions, $\boldsymbol{\theta}^{\pm}$ are the corresponding rotational degrees of freedom, and $\langle \boldsymbol{\varepsilon} \rangle$ is the imposed macroscopic strain tensor. After each finite element simulation, nodal forces $\mathbf{f}_j$ and positions of the boundary nodes $\mathbf{y}_j$ are extracted and post-processed to compute the volume-averaged (macroscopic) stress via,
\begin{align}
    \langle \boldsymbol{\sigma} \rangle = \frac{1}{\lvert\omega\rvert} \sum_{j\in\partial\omega} \mathbf{f}_j \otimes \mathbf{y}_j \,,
    \label{eq:averaged_stress}
\end{align}
where $\lvert\omega\rvert$ denotes the unit-cell volume and $\partial \omega$ is the boundary of the unit cell \cite{Gartner_2021}. The averaged stress and strain are then related linearly through
\begin{align}
    \overline{\langle \boldsymbol{\sigma} \rangle} = \mathbf{C}^{*} \overline{\langle \boldsymbol{\varepsilon} \rangle} \,,
    \label{eq:homogenized_constitutive}
\end{align}
where the overline denotes Voigt notation. The homogenized stiffness $\mathbf{C}^{*} = \mathbf{C}^{*} ( \mathbf{p} ) \in\mathbb{R}^{6\times 6}$ is thus obtained column by column by imposing six independent macroscopic unit strain states,
\begin{align}
    \begin{split}
        \overline{\langle \boldsymbol{\varepsilon} \rangle}_1 &= [1, 0, 0, 0, 0, 0]^{\mathrm{T}} \,,\\
        \overline{\langle \boldsymbol{\varepsilon} \rangle}_2 &= [0, 1, 0, 0, 0, 0]^{\mathrm{T}} \,, \\
        \overline{\langle \boldsymbol{\varepsilon} \rangle}_3 &= [0, 0, 1, 0, 0, 0]^{\mathrm{T}} \,, \\
        \overline{\langle \boldsymbol{\varepsilon} \rangle}_4 &= [0, 0, 0, 1, 0, 0]^{\mathrm{T}} \,, \\
        \overline{\langle \boldsymbol{\varepsilon} \rangle}_5 &= [0, 0, 0, 0, 1, 0]^{\mathrm{T}} \,, \\
        \overline{\langle \boldsymbol{\varepsilon} \rangle}_6 &= [0, 0, 0, 0, 0, 1]^{\mathrm{T}} \,, \\
    \end{split}
\end{align}
whose homogenized stress responses form the six columns of the symmetric matrix $\mathbf{C}^{*}$, which has $21$ independent components.

% ------------------------------------------------------------
\subsection{Cholesky-constrained stiffness surrogate}
\label{subsec:stiffness_surrogate}
% ------------------------------------------------------------

The stiffness surrogate is trained to approximate the mapping from the input vector given in \eqref{eq:stiffness_input} to the homogenized stiffness tensor $\mathbf{C}^{*}$. Direct prediction of stiffness tensor entries can lead to non-physical predictions, especially if the predicted tensor is not positive definite. Since the stiffness tensor is used directly in the finite element equilibrium equation, $\mathbf{C}^{*}$ must be positive definite so that the strain energy density, 
\begin{align}
    \Psi = \frac{1}{2} \overline{ \langle\boldsymbol{\varepsilon}\rangle }^{\mathrm{T}} \mathbf{C}^{*} \overline{\langle\boldsymbol{\varepsilon}\rangle} \,,
\end{align}
remains non-negative for stable and thermodynamically admissible simulations. To address this issue, and following the idea of enforcing physical constraints directly through the network architecture \cite{Linden_2023,Stollberg_2025}, the stiffness tensor is represented using a Cholesky factorization,
\begin{align}
    \mathbf{C}^{*} = \mathbf{G}\mathbf{G}^{\mathrm{T}} \,,
    \label{eq:cholesky_reconstruction}
\end{align}
where $\mathbf{G} = \mathbf{G} ( \mathbf{p} )$ is a lower triangular matrix with positive diagonal entries. Instead of predicting $\mathbf{C}^{*}$ directly, our neural network predicts the independent components of $\mathbf{G}$, such that the stiffness tensor can be reconstructed using Eq.~\eqref{eq:cholesky_reconstruction}. This construction automatically preserves the symmetry and positive definiteness of the predicted stiffness response.

Although $\mathbf{G}$ generally contains 21 independent components (matching the 21 independent entries of the symmetric $6\times6$ stiffness tensor in Voigt notation), the geometric configuration and symmetry of the superimposed lattice family considered here reduce this to six. Specifically, because the FCC component does not include struts on two cell faces normal to one axis (see Figure~\ref{fig:superimposed_lattice}), that axis is distinguished from the other two and the effective material exhibits tetragonal symmetry. Therefore, $\mathbf{G}$ can be written as
\begin{align}
    \mathbf{G} = \begin{bmatrix}
        G_{11} & 0 & 0 & 0 & 0 & 0 \\
        G_{21} & G_{22} & 0 & 0 & 0 & 0 \\
        G_{31} & G_{32} & G_{33} & 0 & 0 & 0 \\
        0 & 0 & 0 & G_{44} & 0 & 0 \\
        0 & 0 & 0 & 0 & G_{55} & 0 \\
        0 & 0 & 0 & 0 & 0 & G_{66}
    \end{bmatrix} \,,
    \label{eq:g_matrix_structure}
\end{align}
with only the six entries $G_{11}$, $G_{22}$, $G_{21}$, $G_{33}$, $G_{55}$, and $G_{44}$ independent. The remaining entries $G_{66}$, $G_{32}$, and $G_{31}$ are recovered from the symmetry relations of the Cholesky representation, given in~\ref{app:symmetry_relations}. This exploitation of symmetry reduces the complexity of the neural network output while ensuring physical consistency of the predicted stiffness matrices.

The stiffness surrogate used in this work, shown in Fig.~\ref{fig:cholesky_nn}, is a feedforward neural network with four hidden layers and 128 neurons in each hidden layer, using the softplus activation function
\begin{align}
    s(x)=\log(1+\exp(x)) \,,
    \label{eq:softplus}
\end{align}
throughout. This architecture was set up using the open-source Flux library available for the programming language Julia \cite{Innes_2018}. The feedforward architecture predicts a latent vector
\begin{align}
    \mathbf{v} = [v_1,v_2,v_3,v_4,v_5,v_6]^{\mathrm{T}} \,,
\end{align}
which is subsequently transformed into the six independent components $G_{11}$, $G_{22}$, $G_{21}$, $G_{33}$, $G_{55}$, and $G_{44}$ through an additional constrained layer that enforces the required symmetry and positivity, again through application of \eqref{eq:softplus}. Consequently, the constrained mapping is written as
\begin{subequations}
    \begin{align}
        G_{11} &= s(v_1) + \sqrt{ s(v_4)^2 + \frac{v_3^2 s(v_4)^2}{s(v_2)^2} } \,, \label{eq:g11_mapping}\\
        G_{22} &= s(v_2) \,, \label{eq:g22_mapping}\\
        G_{21} &= v_3 \,, \label{eq:g21_mapping}\\
        G_{33} &= s(v_4) \,, \label{eq:g33_mapping}\\
        G_{55} &= s(v_5) \,, \label{eq:g55_mapping}\\
        G_{44} &= s(v_6) \,. \label{eq:g44_mapping}
    \end{align}
\end{subequations}
The final stiffness prediction is then obtained from $\mathbf{C}^{*}_{\mathrm{NN}}=\mathbf{G}_{\mathrm{NN}}\mathbf{G}_{\mathrm{NN}}^{\mathrm{T}}$, where the subscript $\mathrm{NN}$ denotes quantities predicted by the trained network, as opposed to the reference (ground-truth) homogenized quantities used for training.

The neural network is trained by minimizing the mean squared error between the predicted and reference Cholesky components of Eq.~\eqref{eq:cholesky_reconstruction}, using a dataset of 1{,}926 homogenized stiffness samples generated by sampling the input parameters over $a_i \in [10^{-4}, 0.5]$, $E \in [0.01, 300]$, and $\nu \in [0, 0.5]$, with $1{,}540$ samples (80\,\%) used for training and 386 (20\,\%) held out for testing. Before training, both input and output quantities are normalized using min--max scaling, since the input variables have different numerical ranges, especially due to the Young's modulus. The network is trained using the AMSGrad optimizer (learning rate $3\times10^{-3}$) for $50{,}000$ epochs.

\begin{figure}[htbp]
    \centering
    \includegraphics[width=\columnwidth]{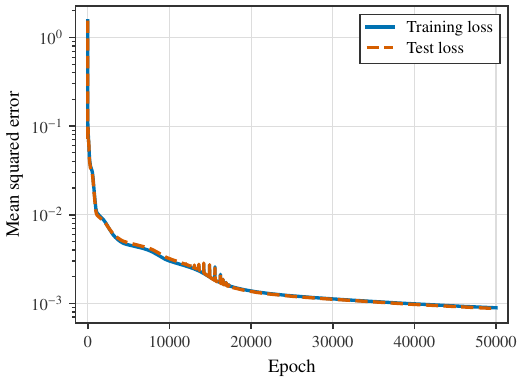}
    \caption{Training and testing loss history of the stiffness surrogate over 50{,}000 epochs (mean squared error, log scale).}
    \label{fig:loss_stiffness}
\end{figure}

Figure~\ref{fig:loss_stiffness} shows the training and testing loss history. Both losses decrease by more than three orders of magnitude over the $50{,}000$ epochs, reaching a final mean squared error of $8.9\times10^{-4}$ on the training set and $8.7\times10^{-4}$ on the test set. The test loss stays marginally below the training loss throughout and the two curves never diverge, which shows that the Cholesky-constrained surrogate generalizes to unseen lattice configurations within the sampled design range rather than overfitting the training data. The steepest reduction occurs during the first few thousand epochs, after which the loss decays slowly toward this plateau, so that the chosen budget of $50{,}000$ epochs is sufficient for the loss to saturate.

\begin{figure*}[htbp]
    \centering
    \includegraphics[width=0.95\textwidth]{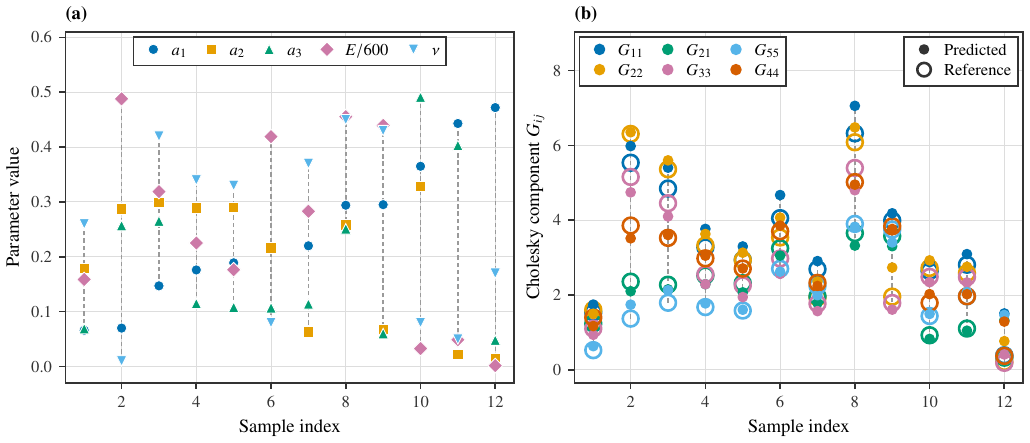}
    \caption{Validation of the stiffness surrogate on selected test samples: (a) input material and lattice parameters used for testing, where the Young's modulus is plotted as $E/600$ so that it shares the axis range of the dimensionless aspect ratios and Poisson's ratio, and (b) predicted (filled markers) and reference (open circles) Cholesky components for the same samples. The close agreement between the two indicates that the surrogate captures the variation of effective stiffness with respect to lattice parameters.}
    \label{fig:stiffness_validation}
\end{figure*}

\begin{figure}[htbp]
    \centering
    \includegraphics[width=\columnwidth]{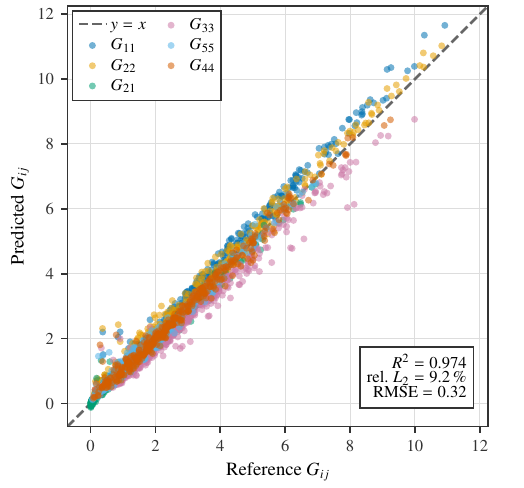}
    \caption{Parity plot of the stiffness surrogate over the complete held-out test set (all six Cholesky components of the 386 test samples). The dashed line marks perfect agreement ($y=x$). The coefficient of determination $R^2$, the relative $L_2$ error, and the RMSE are reported in the panel.}
    \label{fig:parity_stiffness}
\end{figure}

The trained stiffness surrogate is evaluated on the held-out test data to assess its predictive accuracy. Figure~\ref{fig:stiffness_validation} illustrates this for twelve representative test samples: panel~(a) shows the input material and lattice parameters, and panel~(b) compares the predicted and reference Cholesky components. The predicted components agree closely with the reference values across the full range of sampled configurations. Over the complete test set of $386$ samples, the surrogate reproduces the homogenized Cholesky components with a coefficient of determination $R^2 = 0.974$, an aggregated relative $L_2$ error of $9.2\,\%$, and a root-mean-square error (RMSE) of $0.32$, as shown by the parity plot in Fig.~\ref{fig:parity_stiffness}. The RMSE is small relative to the range of the Cholesky components, which extends from $0$ to about $11$.

The accuracy is not uniform across the six components. The shear-related factor $G_{44}$ is reproduced most accurately, with a relative $L_2$ error of $6.3\,\%$ and $R^2 = 0.986$, followed by $G_{22}$ ($7.6\,\%$), the off-diagonal factor $G_{21}$ ($7.7\,\%$), and the shear factor $G_{55}$ ($7.8\,\%$), all with $R^2 \geq 0.979$. The largest errors occur for the diagonal factors $G_{11}$ ($9.4\,\%$, $R^2 = 0.965$) and $G_{33}$ ($13.0\,\%$, $R^2 = 0.941$), which govern the normal stiffness and are the hardest to fit because they combine the largest component magnitudes with the anisotropy introduced by the distinguished tetragonal axis. This ordering is reflected in the parity plot of Fig.~\ref{fig:parity_stiffness}, where the residual scatter is concentrated in the normal-stiffness components and grows toward the upper end of the value range, while all components remain tightly clustered around the line of perfect agreement.

% ------------------------------------------------------------
\subsection{Relative density surrogate}
\label{subsec:density_surrogate}
% ------------------------------------------------------------

The relative density $\tilde{\rho}$ of a lattice unit cell is required for imposing the material constraint during microscale optimization. A separate neural network is therefore trained to predict the relative density from the geometric lattice parameter vector $\mathbf{a}$ of Eq.~\eqref{eq:lattice_parameters},
\begin{align}
    \mathbf{a} \longmapsto \tilde{\rho}_{\mathrm{NN}} \,.
    \label{eq:density_mapping}
\end{align}

For the superimposed lattice, a simple volume-ratio approximation can overestimate the true solid volume because struts from different lattice components may overlap, and this overestimation grows as the aspect ratios increase. Relative density is therefore estimated using Monte Carlo sampling instead, which was shown to be more accurate than the direct volume-ratio estimate in the comparison of density-estimation methods \cite{Stollberg_2025,Souza_2018}. For that, a set of $N_\mathrm{s}$ points with microscale coordinates $\mathbf{y}_i$ is sampled inside the unit cell. For each point, an indicator function is evaluated:
\begin{align}
    I(\mathbf{y}_i) = \begin{cases}
        1, & \text{if } \mathbf{y}_i \text{ lies inside the solid lattice region},\\
        0, & \text{otherwise}.
    \end{cases}
    \label{eq:indicator_function}
\end{align}
The relative density is then approximated as
\begin{align}
    \tilde{\rho} \approx \frac{1}{N_{\mathrm{s}}} \sum_{i=1}^{N_\mathrm{s}} I(\mathbf{y}_i) \,.
    \label{eq:monte_carlo_density}
\end{align}
The number of sample points $N_{\mathrm{s}}$ is chosen adaptively: at least $500$ points are used, and further points are then added until either a total of $2{,}000$ points is reached or the relative change of the estimated density between two successive iterations falls below $1 \times 10^{-3}$.

\begin{figure}[htbp]
    \centering
    \includegraphics[width=\columnwidth]{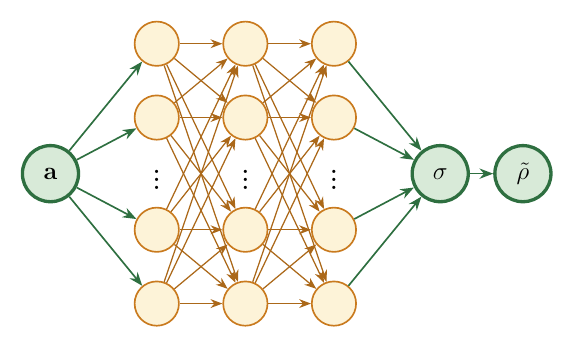}
    \caption{Relative density surrogate architecture: the network maps the aspect-ratio vector $\mathbf{a}=[a_1,a_2,a_3]^{\mathrm{T}}$ through three hidden layers to a sigmoid output $\sigma$, giving the predicted relative density $\tilde{\rho}$.}
    \label{fig:density_nn}
\end{figure}

The density surrogate, shown in Fig.~\ref{fig:density_nn}, is a feedforward neural network with three hidden layers and 64 neurons in each hidden layer, using ReLU activation in every hidden layer. It maps $\mathbf{a}$ to the predicted relative density $\tilde{\rho}_{\mathrm{NN}}$ through a sigmoid output activation $\sigma$ that keeps the prediction within the physical range $(0,1)$. It is trained using mean squared error on a dataset of $1{,}500$ lattice configurations, with $1{,}200$ configurations (80\,\%) used for training and 300 (20\,\%) held out for testing. Because the relative density depends only on the lattice geometry through the Monte Carlo estimate, and the resulting mapping is simpler than the stiffness mapping, a smaller dataset and a simpler network than those of the stiffness surrogate are sufficient. The network is trained using the Adam optimizer (learning rate $1\times10^{-3}$) for 500 epochs.

\begin{figure}[htbp]
    \centering
    \includegraphics[width=\columnwidth]{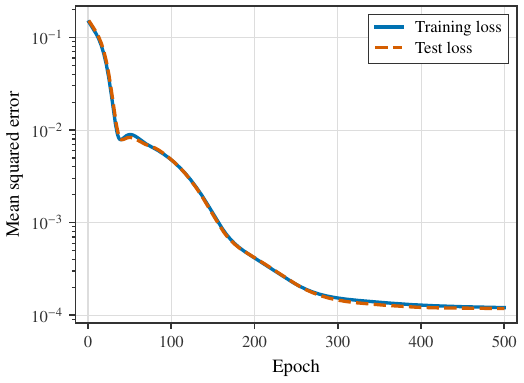}
    \caption{Training and testing loss history of the density surrogate over 500 epochs (mean squared error, log scale).}
    \label{fig:loss_density}
\end{figure}

Figure~\ref{fig:loss_density} shows the training and testing loss history. Both losses drop by three orders of magnitude over the $500$ epochs, reaching a final mean squared error of $1.2\times10^{-4}$ on both the training and test sets. The two curves track each other throughout and the test loss stays at or below the training loss, so the density surrogate also generalizes without overfitting. The loss converges within the $500$ epochs.

\begin{figure*}[tp]
    \centering
    \includegraphics[width=0.95\textwidth]{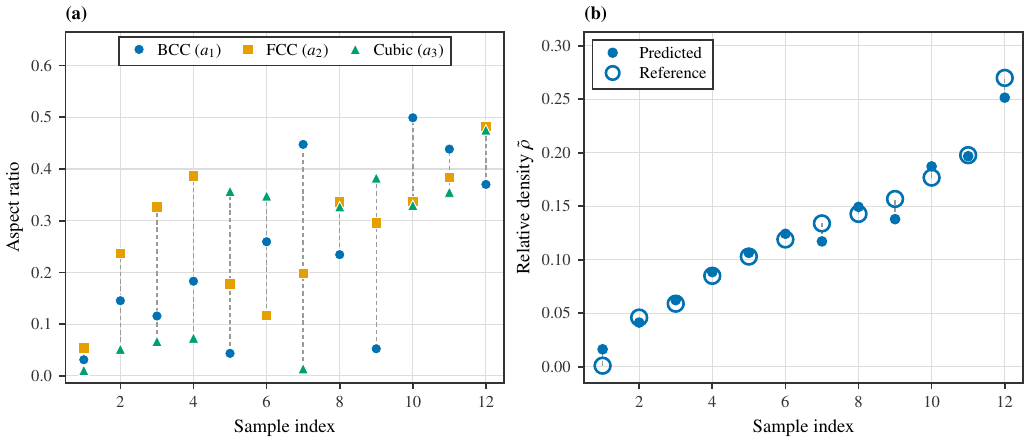}
    \caption{Validation of the relative density surrogate on selected test samples: (a) input geometric parameters for the BCC, FCC, and simple cubic lattice components, and (b) predicted (filled markers) and Monte Carlo-estimated (open circles) relative densities.}
    \label{fig:density_validation}
\end{figure*}

\begin{figure}[htbp]
    \centering
    \includegraphics[width=\columnwidth]{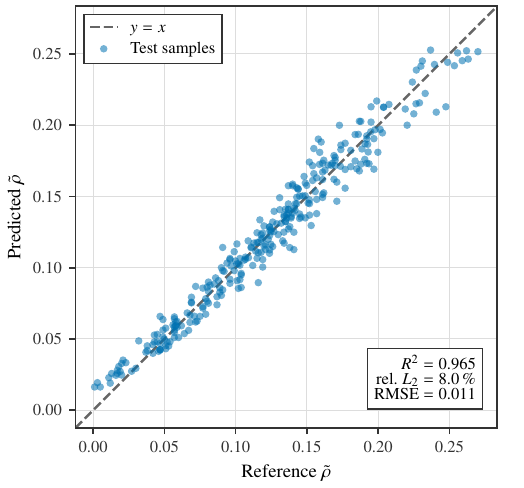}
    \caption{Parity plot of the relative density surrogate over the 300 held-out test samples. The dashed line marks perfect agreement ($y=x$). The coefficient of determination $R^2$, the relative $L_2$ error, and the RMSE are reported in the panel.}
    \label{fig:parity_density}
\end{figure}

The trained density surrogate is evaluated on the held-out test data to assess its predictive accuracy. Figure~\ref{fig:density_validation} illustrates the agreement for twelve representative samples spanning the full density range: panel~(a) reports the input geometric parameters of the BCC, FCC, and simple cubic components, and panel~(b) compares the predicted relative density with the corresponding Monte Carlo estimate, which the predictions follow closely. Over the complete test set of 300 samples, the surrogate reproduces the Monte Carlo densities with a coefficient of determination $R^2 = 0.965$ and an aggregated relative $L_2$ error of $8.0\,\%$, as shown by the parity plot in Fig.~\ref{fig:parity_density}. The predictions cluster tightly around the line of perfect agreement across the full range of sampled densities. The largest relative deviations occur at the very lowest densities, where a small absolute error corresponds to a comparatively large relative one. These points dominate the aggregated relative $L_2$ error, whereas the absolute error remains small, with a root-mean-square error of $1.1\times10^{-2}$.

In summary, the microscale behavior of the superimposed lattice is represented by two complementary neural network surrogates: a Cholesky-constrained stiffness surrogate that returns a physically admissible homogenized stiffness tensor, and a density surrogate that returns the corresponding relative density. Both are trained offline on homogenization data and, once trained, replace repeated online homogenization within the optimization loop. Their inputs, outputs, architectures, and dataset sizes are summarized in Table~\ref{tab:surrogate_models}.

\begin{table}[t]
    \centering
    \footnotesize
    \setlength{\tabcolsep}{4pt}
    \begin{tabularx}{\columnwidth}{@{}l c c >{\raggedright\arraybackslash}X c@{}}
        \toprule
        Model & Input & Output & Architecture and activation & Size \\
        \midrule
        Stiffness
        & $\mathbf{p}$ & $\mathbf{G}$ & 4 hidden layers, 128 neurons each; softplus & 1{,}926 \\
        Density & $\mathbf{a}$ & $\tilde{\rho}$ & 3 hidden layers, 64 neurons each; ReLU (hidden), sigmoid (output) & 1{,}500 \\
        \bottomrule
    \end{tabularx}
    \caption{Summary of the neural network surrogate models used in the proposed framework. Inputs $\mathbf{p}$ and $\mathbf{a}$ are defined in Eqs.~\eqref{eq:stiffness_input} and~\eqref{eq:lattice_parameters}, respectively.}
    \label{tab:surrogate_models}
\end{table}

% ============================================================
\section{Two-stage multiscale topology optimization}
\label{sec:optimization}
% ============================================================

Building on the machine-learning surrogates of Section~\ref{sec:surrogates}, this section formulates the two-stage multiscale topology optimization \cite{Zhang_2021}, in which the trained stiffness and density surrogates replace repeated online unit-cell homogenization inside the optimization loop. The first optimization stage determines the macroscale material layout using the SIMP method \cite{Bendsoe_2004,Sigmund_2013}. The second stage optimizes the local lattice parameters within the solid region obtained from the first stage.

In the macroscale stage, the structure occupies a computational domain $\Omega$ that is discretized into $n_{\mathrm{el}}$ finite elements. Following the SIMP approach, each element $e$ is assigned a pseudo-density variable $\gamma_e$ (a topological design variable, not a physical relative density), and these variables are collected in the design vector $\boldsymbol{\gamma} = [\gamma_1,\dots,\gamma_{n_{\mathrm{el}}}]^{\mathrm{T}}$. Each design variable is bounded as
\begin{align}
    \gamma_{\min} \leq \gamma_e \leq 1 \,,
\end{align}
where the lower bound $\gamma_{\min} \ll 1$ prevents the element stiffness matrices from becoming singular. The penalized element stiffness matrix $\mathbf{K}^{\mathrm{mac}}_{e}$ at this macroscale (SIMP) stage is written as
\begin{align}
    \mathbf{K}^{\mathrm{mac}}_{e}(\gamma_e) = \gamma_e^{q}\,\mathbf{K}^{0}_{e} \,,
    \label{eq:simp_stiffness}
\end{align}
where $q$ is the penalization factor, and $\mathbf{K}^{0}_{e}$ is the element stiffness matrix computed for the element filled with full base material. The macroscale compliance minimization problem is
\begin{align}
\begin{aligned}
    \min_{\boldsymbol{\gamma}} \quad
    & J_{\mathrm{mac}}(\boldsymbol{\gamma}) = \mathbf{f}^{\mathrm{T}}\mathbf{u} \,, \\
    \text{subject to} \quad
    & \mathbf{K}^{\mathrm{mac}}(\boldsymbol{\gamma})\mathbf{u} = \mathbf{f} \,, \\
    & \sum_{e=1}^{n_\mathrm{el}} \gamma_e V_e \leq V_{\max} \,, \\
    & \gamma_{\min} \leq \gamma_e \leq 1 \,,
\end{aligned}
\label{eq:simp_problem}
\end{align}
where $\mathbf{u}$ is the global displacement vector obtained from the equilibrium equation, $\mathbf{f}$ is the applied external load vector, $\mathbf{K}^{\mathrm{mac}}$ is the assembled global stiffness matrix, $V_e$ is the volume of element $e$, and $V_{\max}$ is the prescribed upper bound on the total macroscale volume.

The macroscale density variables are updated using the optimality-criteria (OC) method. This update requires the sensitivity of the compliance with respect to each design variable, which is obtained using the adjoint method \cite{Bendsoe_2004}. Since the compliance objective is self-adjoint, the sensitivity takes the closed form
\begin{align}
    \frac{\partial J_{\mathrm{mac}}}{\partial \gamma_e} = -\,\mathbf{u}_e^{\mathrm{T}} \frac{\partial \mathbf{K}^{\mathrm{mac}}_{e}}{\partial \gamma_e}\, \mathbf{u}_e = -\,q\,\gamma_e^{q-1}\,\mathbf{u}_e^{\mathrm{T}} \mathbf{K}^{0}_{e}\, \mathbf{u}_e \,,
    \label{eq:compliance_sensitivity}
\end{align}
with the element displacement vector $\mathbf{u}_e$. Because $\mathbf{K}^{0}_{e}$ is positive semidefinite, this sensitivity is non-positive, so increasing $\gamma_e$ never increases the compliance. To prevent numerical instabilities such as checkerboarding and mesh-dependent patterns, a linear (cone-weighted) sensitivity filter is applied to these sensitivities before each update \cite{Sigmund_1998}:
\begin{align}
    \begin{aligned}
        \left(\frac{\partial J_{\mathrm{mac}}}{\partial \gamma_e}\right)_{\mathrm{filtered}} &= \frac{1}{\gamma_e \sum_{j\in \mathcal{N}_e} w_{ej}} \sum_{j\in \mathcal{N}_e} w_{ej}\,\gamma_j\,\frac{\partial J_{\mathrm{mac}}}{\partial \gamma_j} \,, \qquad \\
        w_{ej} &= \max(0,\, r_{\min} - \mathrm{dist}(e,j)) \,,
    \end{aligned}
    \label{eq:sensitivity_filter}
\end{align}
where $\mathcal{N}_e$ is the set of elements within a filter radius $r_{\min}$ of element $e$, and the linearly decaying weights $w_{ej}$ give closer neighbors more influence in the filtered sensitivity, reducing the tendency to form single-element-wide, mesh-dependent members. The filtered sensitivities are then used to update the pseudo-densities through the classical optimality-criteria bisection scheme~\cite{Bendsoe_2004,Sigmund_2001}:
\begin{align}
\begin{split}
    \gamma_e^{(k+1)} ={}& \max\big(\gamma_{\min}, \\
    &\quad \max(\gamma_e^{(k)}-m, \\
    &\quad \min(1,\, \\
    &\quad\min(\gamma_e^{(k)}+m,\, \gamma_e^{(k)} B_e^{\eta})))\big) \,,
\end{split}
\label{eq:oc_update}
\end{align}
with
\begin{align}
    B_e = -\frac{1}{\Lambda\, V_e} \left( \frac{\partial J_{\mathrm{mac}}}{\partial \gamma_e} \right)^{(k)}_{\mathrm{filtered}} \,,
    \label{eq:oc_be}
\end{align}
where $m$ denotes the move limit bounding the change in $\gamma_e$ per iteration, $\eta$ is a damping exponent, and $\Lambda$ is a Lagrange multiplier updated by bisection at each iteration to satisfy the volume constraint of Eq.~\eqref{eq:simp_problem}.

The macroscale optimization is terminated once all constraints are satisfied and the compliance is stable over a window of $N$ successive iterations, following the convergence criterion
\begin{align}
    \frac{\sum_{i=1}^{N}\left| J_{\mathrm{mac}}^{(k-i+1)} - J_{\mathrm{mac}}^{(k-N-i+1)}\right|}{\sum_{i=1}^{N} J_{\mathrm{mac}}^{(k-i+1)}} \leq \epsilon \ll 1 \,,
    \label{eq:macro_convergence}
\end{align}
where $k$ is the current design iteration, $N$ is the length of the averaging window, and $\epsilon$ is a prescribed tolerance \cite{Huang_2007}.

After the macroscale optimization converges, the density field is thresholded to identify the solid region $\Omega_s$. Microscale optimization is then performed only in this region. Each solid element is assigned a local lattice parameter vector
\begin{align}
    \mathbf{a}_e = [a_{1e},a_{2e},a_{3e}]^{\mathrm{T}} \,,
\end{align}
which enters the stiffness surrogate through the local input vector $\mathbf{p}_e$ of Eq.~\eqref{eq:stiffness_input}, with the aspect ratios set to $\mathbf{a}_e$, while $E$ and $\nu$ are fixed at the base-material values. The element stiffness matrix $\mathbf{K}^{\mathrm{mic}}_{e}$ at this microscale (lattice) stage is computed from the neural network-predicted homogenized stiffness tensor,
\begin{align}
    \mathbf{K}^{\mathrm{mic}}_{e}(\mathbf{p}_e) = \int_{\Omega_e} \mathbf{B}_e^{\mathrm{T}} \mathbf{C}^{*}_{\mathrm{NN}}(\mathbf{p}_e) \mathbf{B}_e \,\mathrm{d}\Omega_{e} \,,
    \label{eq:element_stiffness_nn}
\end{align}
$\mathbf{B}_e$ being the strain-displacement matrix.

The microscale optimization problem is written as
\begin{align}
\begin{aligned}
    \min_{\{\mathbf{a}_e\}} \quad
    & J_{\mathrm{mic}}(\{\mathbf{a}_e\}) = \mathbf{f}^{\mathrm{T}}\mathbf{u} \,, \\
    \text{subject to} \quad
    & \mathbf{K}^{\mathrm{mic}}(\{\mathbf{a}_e\})\mathbf{u} = \mathbf{f} \,, \\
    & \sum_{e\in\Omega_s}
    \tilde{\rho}_{\mathrm{NN}}(\mathbf{a}_e) V_e
    \leq V_{\mathrm{micro}} \,, \\
    & a_i^{\min} \leq a_{ie} \leq a_i^{\max} \,,
    \qquad i=1,2,3 \,.
\end{aligned}
\label{eq:micro_problem}
\end{align}
The method of moving asymptotes \cite{Svanberg_1987,Svanberg_2002} is used to update the lattice parameters. Since both $\mathbf{C}^{*}_{\mathrm{NN}}$ and $\tilde{\rho}_{\mathrm{NN}}$ are differentiable neural network models, the required sensitivities can be computed efficiently through forward-mode automatic differentiation \cite{Revels_2016}. At each design iteration, the stiffness surrogate predicts the homogenized stiffness tensor for each solid element and the density surrogate predicts the corresponding relative density. These predictions are used to assemble the global stiffness matrix, solve the finite element equilibrium equation, evaluate compliance, and enforce the material constraint. The microscale optimization is terminated when either the absolute change of the design variables or the relative change of the objective between two successive design iterations falls below the tolerance $\epsilon$,
\begin{align}
    \max_{e,\,i}\left| a_{ie}^{(k)} - a_{ie}^{(k-1)}\right| < \epsilon \quad\text{or}\quad \frac{\left| J_{\mathrm{mic}}^{(k)} - J_{\mathrm{mic}}^{(k-1)}\right|}{\left| J_{\mathrm{mic}}^{(k)}\right|} < \epsilon \,,
    \label{eq:micro_convergence}
\end{align}
where the maximum in the first criterion is taken over all solid elements $e$ and the three lattice parameters $i=1,2,3$.

% ============================================================
\section{Computational results}
\label{sec:results}
% ============================================================

This section reports the computational results obtained with the proposed framework, using a three-dimensional Messerschmitt-B\"olkow-Blohm (MBB) beam benchmark as a demonstration. For simplicity, all quantities reported in this section are assumed to be dimensionless. The benchmark problem and the macroscale topology optimization are presented first, followed by the microscale lattice optimization, the resulting optimized lattice parameter and density fields, and a closing discussion.

% ------------------------------------------------------------
\subsection{Macroscale topology optimization}
\label{subsec:benchmark_problem}
\begin{figure}[htbp]
    \centering
    \includegraphics[width=\columnwidth]{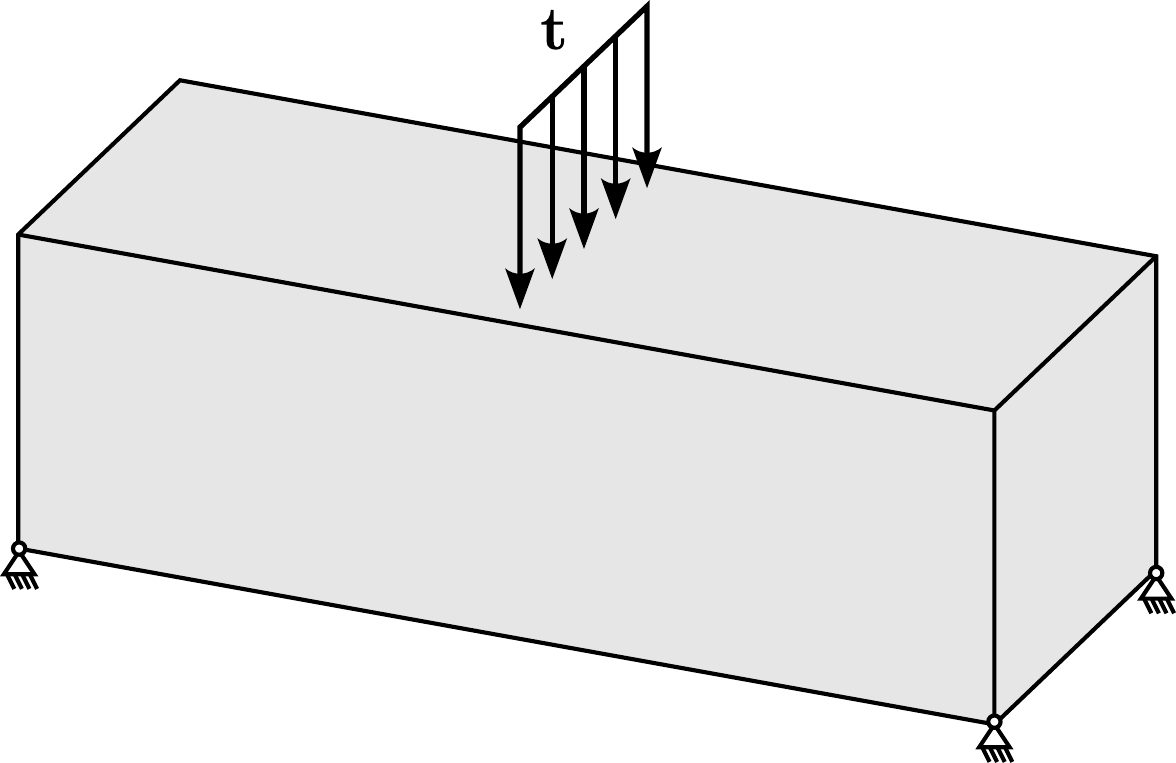}
    \caption{Three-dimensional MBB beam benchmark setup: a downward line load $\mathbf{t}$ is applied along the full depth of the beam at mid-span, perpendicular to the beam length, while the four lower corners are hinged (pinned) supports.}
    \label{fig:mbb_setup}
\end{figure}

% ------------------------------------------------------------
The MBB beam benchmark is a standard test case in topology optimization because it produces a bending-dominated structural response and clear load paths between the loading region and the supports. The beam is modeled as a rectangular three-dimensional design domain of total volume $V_{\mathrm{domain}} = 20\times5\times5$ (length $\times$ height $\times$ depth), discretized into $60\times15\times15 = 13{,}500$ first-order hexahedral finite elements. A downward unit line load $\mathbf{t}$ is applied along the full depth of the beam at the top mid-span, while the four lower corners of the domain are hinged (pinned) supports. The solid base material is assumed to be isotropic and linearly elastic with Young's modulus $E = 45$ and Poisson's ratio $\nu = 0.3$. Figure~\ref{fig:mbb_setup} shows the resulting benchmark setup.

In the first (macroscale) stage, the SIMP problem of Eq.~\eqref{eq:simp_problem} is solved for this benchmark using the optimality-criteria method and sensitivity filter described in Section~\ref{sec:optimization}. The numerical parameters used at this stage are summarized in Table~\ref{tab:macro_params}.

\begin{table}[htbp]
    \centering
    \begin{tabular}{l c c}
        \toprule
        Parameter & Symbol & Value \\
        \midrule
        Penalization factor & $q$ & $3$ \\
        Volume fraction & $V_{\max}/V_{\mathrm{domain}}$ & $0.5$ \\
        Initial pseudo-density & $\gamma_e^{(0)}$ & $0.5$ \\
        Move limit & $m$ & $0.05$ \\
        Damping exponent & $\eta$ & $0.5$ \\
        Filter radius & $r_{\min}$ & $0.5$ \\
        Convergence window & $N$ & $5$ \\
        Convergence tolerance & $\epsilon$ & $10^{-3}$ \\
        \bottomrule
    \end{tabular}
    \caption{Numerical parameters of the macroscale (SIMP) topology optimization.}
    \label{tab:macro_params}
\end{table}

\begin{figure}[htbp]
    \centering
    \includegraphics[width=\columnwidth]{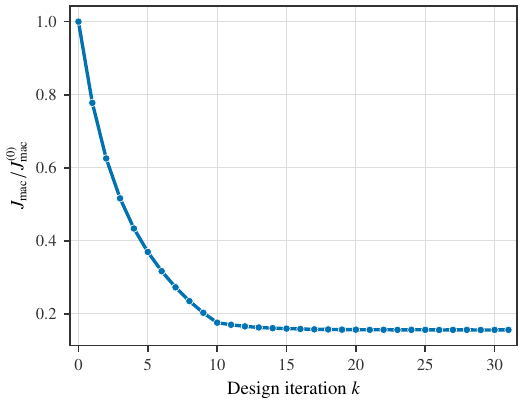}
    \caption{Convergence history of the macroscale (SIMP) topology optimization: the relative compliance $J_{\mathrm{mac}}/J_{\mathrm{mac}}^{(0)}$ decreases to $15.6\,\%$ of its initial value. The volume constraint remains satisfied throughout and is not shown.}
    \label{fig:macro_convergence}
\end{figure}

The convergence history of the macroscale optimization is shown in Fig.~\ref{fig:macro_convergence}. The relative compliance $J_{\mathrm{mac}}/J_{\mathrm{mac}}^{(0)}$ decreases monotonically to $15.6\,\%$ of its initial value. After $12$ iterations it is already very close to the final value, and the remaining iterations only refine the material layout. The volume constraint is satisfied at every iteration as it is enforced automatically through the Lagrange multiplier in \eqref{eq:oc_be}.

After convergence of the macroscale optimization, the resulting density field is thresholded to define the solid region $\Omega_s$ introduced in Section~\ref{sec:optimization}: elements with $\gamma_e \geq 0.5$ are retained as solid ($\gamma_e \leftarrow 1$), while elements with $\gamma_e < 0.5$ are treated as void ($\gamma_e \leftarrow \gamma_{\min} = 1\times10^{-10}$). The microscale optimization is then performed only inside this retained solid region. This decoupled approach reduces the design space of the second stage and avoids optimizing lattice parameters in void regions.

% ------------------------------------------------------------
\subsection{Microscale lattice optimization}
\label{subsec:micro_optimization}
% ------------------------------------------------------------

In the second stage, each solid element obtained from the thresholded SIMP result is assigned a local superimposed lattice unit cell, whose design variables are the local lattice parameters $\mathbf{a}_e=[a_{1e},a_{2e},a_{3e}]^{\mathrm{T}}$ introduced in Section~\ref{sec:optimization}. The lattice parameters are uniformly initialized at $a_{1e}^{(0)}=a_{2e}^{(0)}=a_{3e}^{(0)}=0.4$ and bounded as
\begin{align}
    0.1 \leq a_{ie} \leq 0.4 \,, \qquad i=1,2,3 \,,
    \label{eq:lattice_bounds}
\end{align}
subject to a target volume $V_{\mathrm{micro}} = 0.05 \times V_{\mathrm{domain}}$. These bounds are assumed to correspond to the manufacturable range of strut diameters for the lattice structure of interest. During the microscale optimization, the aspect ratios $a_{1e},a_{2e},a_{3e}$ are the only design variables, while $E$ and $\nu$ are held at the base-material values of Section~\ref{subsec:benchmark_problem}. As in the macroscale stage, the design sensitivities are smoothed by the sensitivity filter of Eq.~\eqref{eq:sensitivity_filter} (radius $r_{\min} = 0.5$). In addition, they are averaged with those of the previous iteration to stabilize the updates \cite{Huang_2007}. The convergence criterion of Eq.~\eqref{eq:micro_convergence} uses the tolerance $\epsilon = 10^{-3}$.

\begin{figure}[htbp]
    \centering
    \includegraphics[width=\columnwidth]{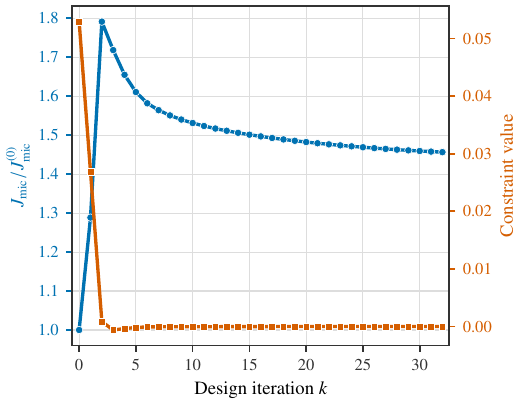}
    \caption{Convergence history of the microscale lattice optimization: relative objective $J_{\mathrm{mic}}/J_{\mathrm{mic}}^{(0)}$ (left axis) and material-constraint value (right axis) over the design iterations.}
    \label{fig:mbb_convergence}
\end{figure}

The microscale optimization is solved with an existing open-source implementation of the method of moving asymptotes as described in Section~\ref{sec:optimization} \cite{Tarek_2023}. It employs the original variant of the method \cite{Svanberg_1987}, in which a single convex separable subproblem is solved at each design iteration. The moving-asymptote update is controlled by three parameters: the initial asymptotes are placed at a distance of $s_{\mathrm{init}} = 0.3$ times the range of each design variable from its current value. When a variable oscillates between two successive iterations, the corresponding asymptotes are moved closer by the factor $s_{\mathrm{decr}} = 0.5$, and when a variable changes monotonically, the asymptote distance is left unchanged, $s_{\mathrm{incr}} = 1.0$. Its convergence history is shown in Fig.~\ref{fig:mbb_convergence}, reporting the relative objective $J_{\mathrm{mic}}/J_{\mathrm{mic}}^{(0)}$ (left axis) and the material-constraint value (right axis) over the design iterations. The initial design ($k=0$) violates the material constraint, with a constraint value of $0.053$. The method of moving asymptotes restores feasibility within two iterations: the constraint value drops to $8\times10^{-4}$ at $k=2$ and becomes negative (strictly feasible) from $k=3$ onward. During this feasibility-restoration phase the compliance increases and reaches a peak of $1.79$ times the initial value at $k=2$, because the initial design assigns the maximum lattice thickness to every solid element and is therefore artificially stiff but over-material. Once the constraint is satisfied, the compliance decreases monotonically as the lattice parameters are redistributed, and reaches a final value of $1.46$ times the compliance of the initial over-material design. The optimization terminates after $32$ iterations, when the convergence criterion of Eq.~\eqref{eq:micro_convergence} is met. The history shows no oscillations, consistent with the smooth, differentiable responses of the surrogates and the conservative asymptote update used in this stage.

\begin{figure*}[htbp]
    \centering
    \begin{subfigure}[b]{0.49\textwidth}
        \centering
        \includegraphics[width=\textwidth]{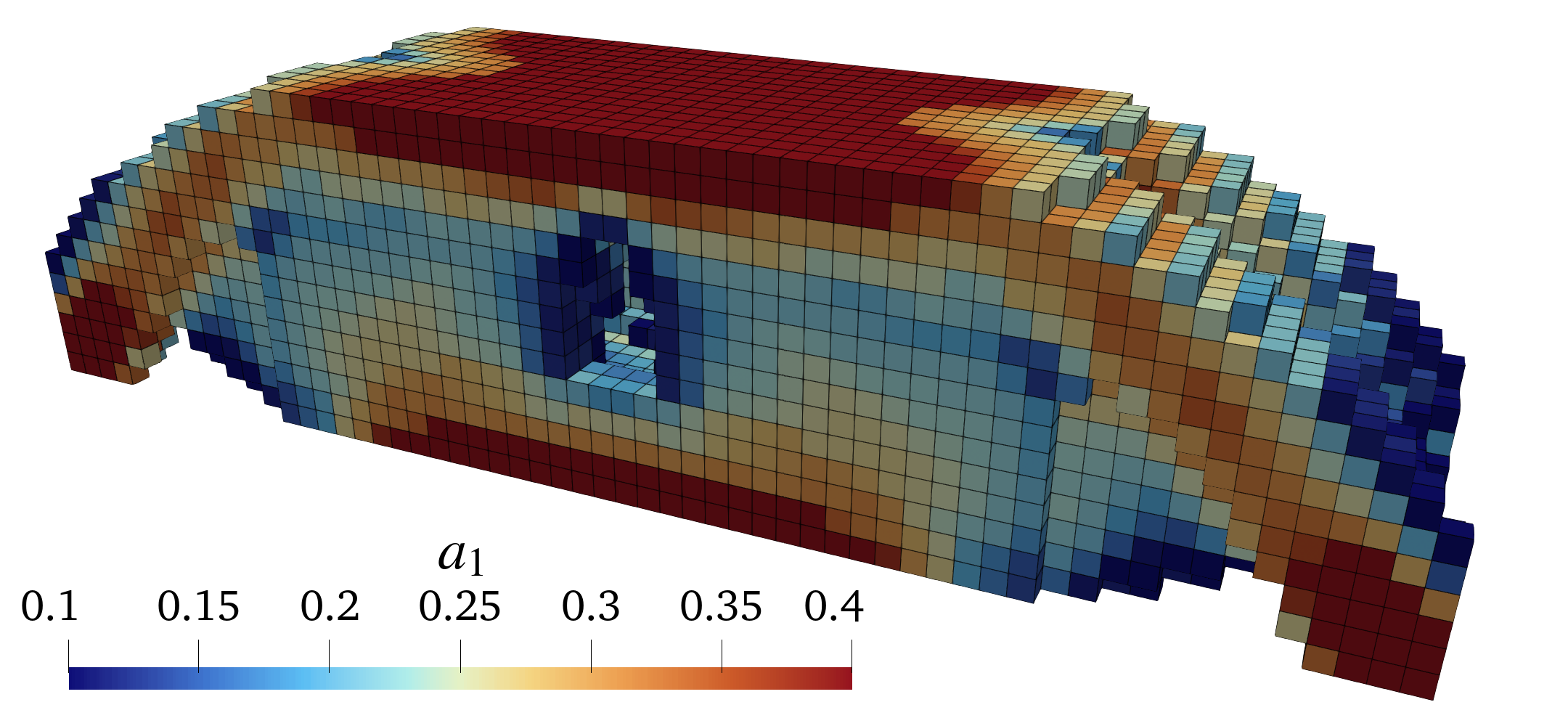}
        \subcaption{Aspect ratio $a_1$ (BCC component)}
    \end{subfigure}
    \hspace{0.01\textwidth}
    \begin{subfigure}[b]{0.49\textwidth}
        \centering
        \includegraphics[width=\textwidth]{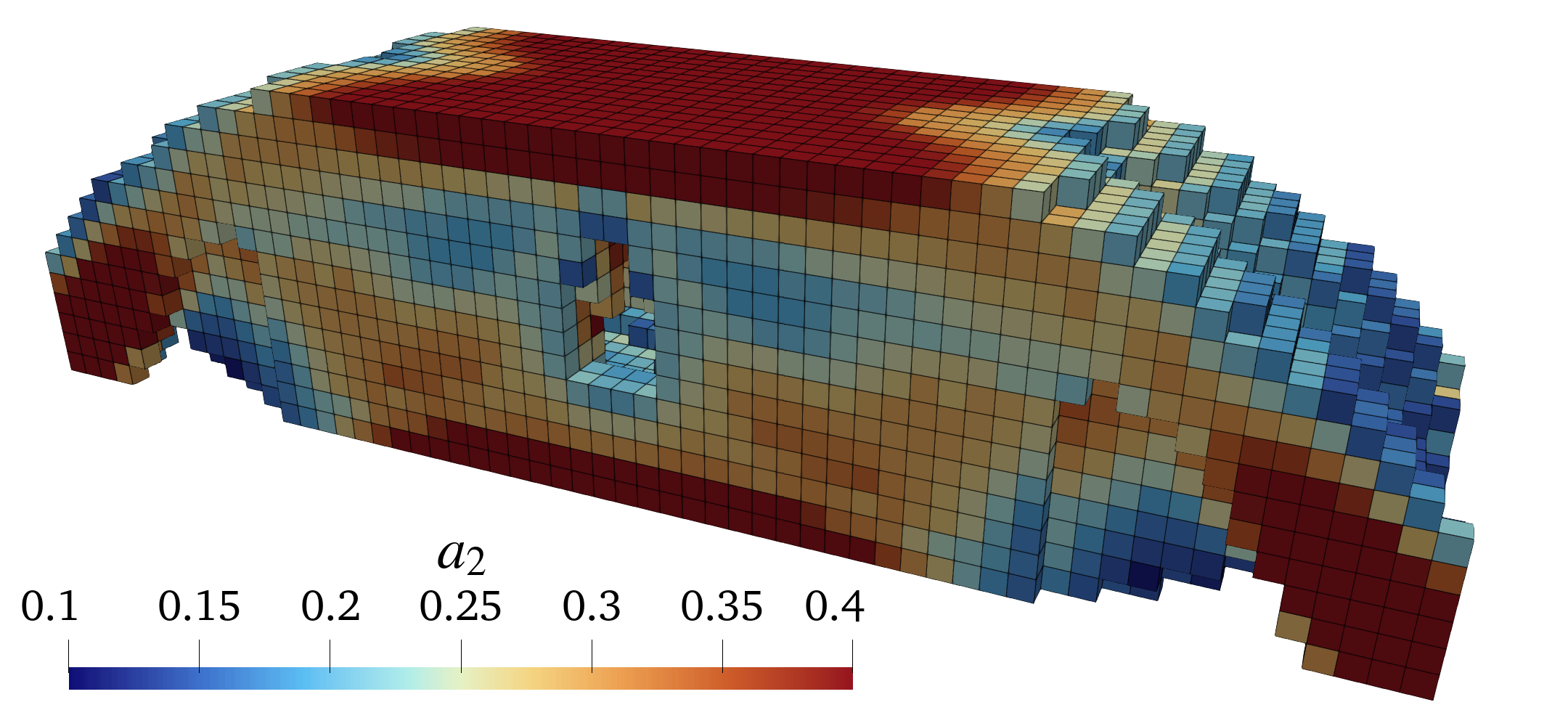}
        \subcaption{Aspect ratio $a_2$ (FCC component)}
    \end{subfigure}

    \vspace{2mm}

    \begin{subfigure}[b]{0.49\textwidth}
        \centering
        \includegraphics[width=\textwidth]{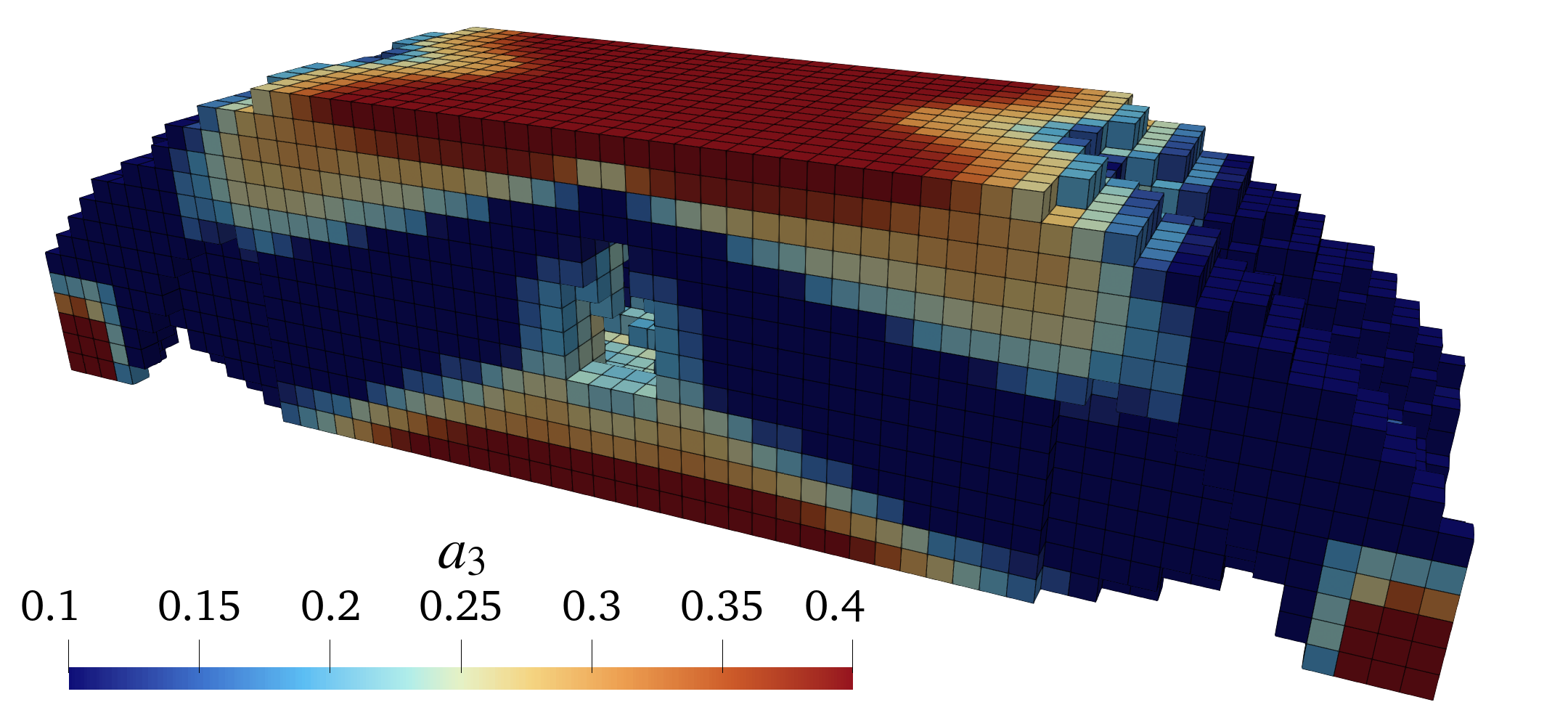}
        \subcaption{Aspect ratio $a_3$ (simple cubic component)}
    \end{subfigure}
    \hspace{0.01\textwidth}
    \begin{subfigure}[b]{0.49\textwidth}
        \centering
        \includegraphics[width=\textwidth]{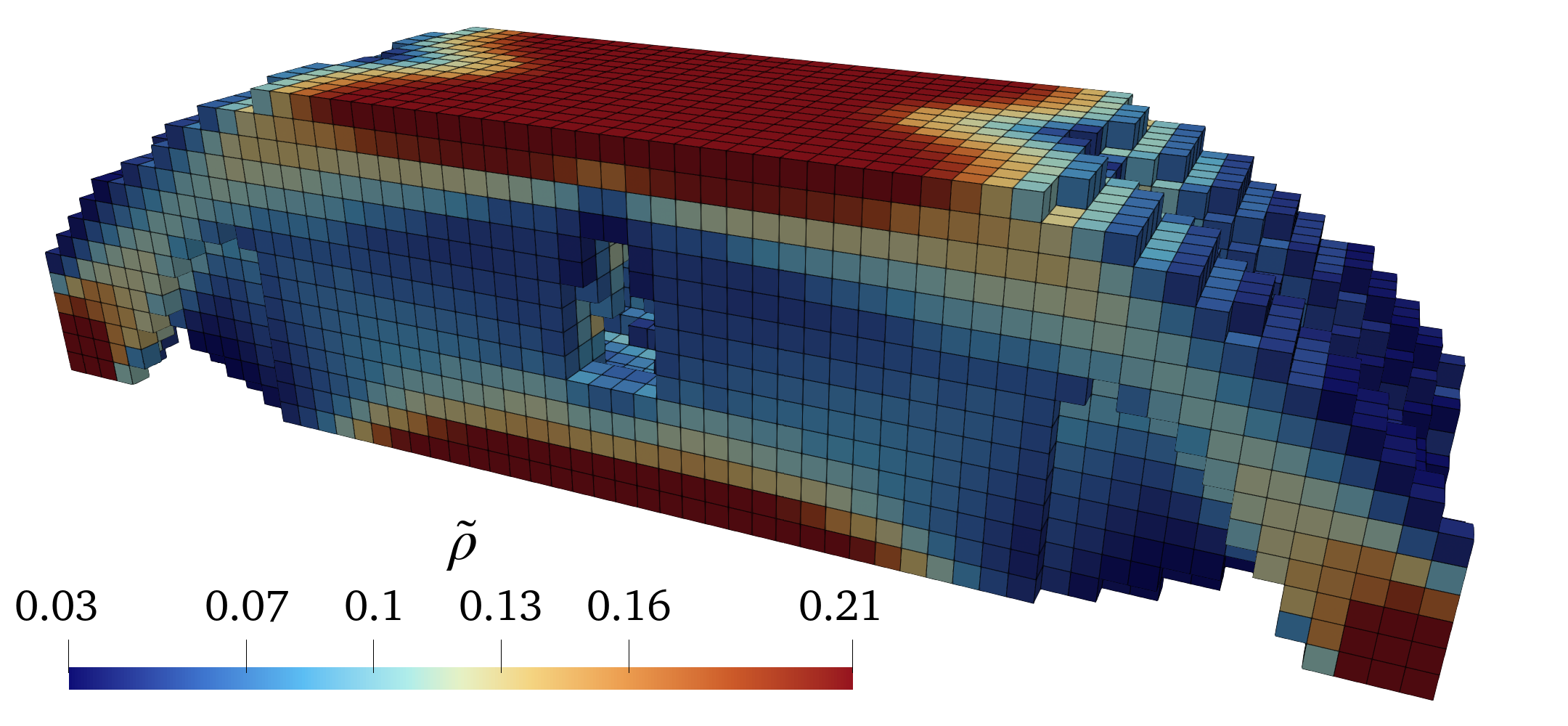}
        \subcaption{Predicted relative density $\tilde{\rho}$}
    \end{subfigure}
    \caption{Optimized spatial fields obtained from microscale lattice optimization: (a) aspect ratio $a_1$ for the BCC component, (b) aspect ratio $a_2$ for the FCC component, (c) aspect ratio $a_3$ for the simple cubic component, and (d) predicted relative density $\tilde{\rho}$.}
    \label{fig:optimized_fields}
\end{figure*}

% ------------------------------------------------------------
\subsection{Optimized lattice parameter and density fields}
\label{subsec:optimized_fields}
% ------------------------------------------------------------

The optimized spatial distributions of $a_1$, $a_2$, $a_3$, and $\tilde{\rho}$ are shown in Fig.~\ref{fig:optimized_fields}. These fields are obtained after microscale optimization and are visualized over the solid region $\Omega_{\mathrm{s}}$ obtained from the thresholded macroscale topology.

The optimized aspect-ratio fields exhibit a clear spatial grading that mirrors the bending-dominated response of the beam. All three components attain their largest values at the upper bound $a_i = 0.4$ in the top and bottom regions of the structure, where the bending-induced normal stresses are largest, and become thinner toward the central region near the neutral axis. The three families are, however, not graded identically: the BCC ($a_1$) and simple cubic ($a_3$) components are strongly concentrated in the top and bottom flanges and thin out to the lower bound $a_i = 0.1$ in the web, whereas the FCC ($a_2$) component remains comparatively large throughout, including near the neutral axis. This is consistent with the differing strut orientations of the three families: the face-diagonal FCC struts contribute shear stiffness and therefore remain useful in the web, while the BCC and simple cubic struts mainly reinforce the flanges against normal stresses. The optimizer thus tunes the local stiffness tensor by adjusting the relative contribution of each lattice family, rather than selecting from a small set of discrete unit-cell types, which illustrates the advantage of the superimposed lattice design space.

The predicted relative density field $\tilde{\rho}$ follows the same trend, ranging from $0.21$ in the top and bottom flanges down to $0.03$ in the central web, so that material is concentrated where the bending stresses are highest and reduced near the neutral axis. This distribution is consistent with compliance minimization under the prescribed material constraint and with the expected behavior of an MBB beam, in which material is mainly retained along the top and bottom load-transfer paths and near the supports.

\begin{figure*}[htbp]
    \centering
    \includegraphics[width=0.95\textwidth]{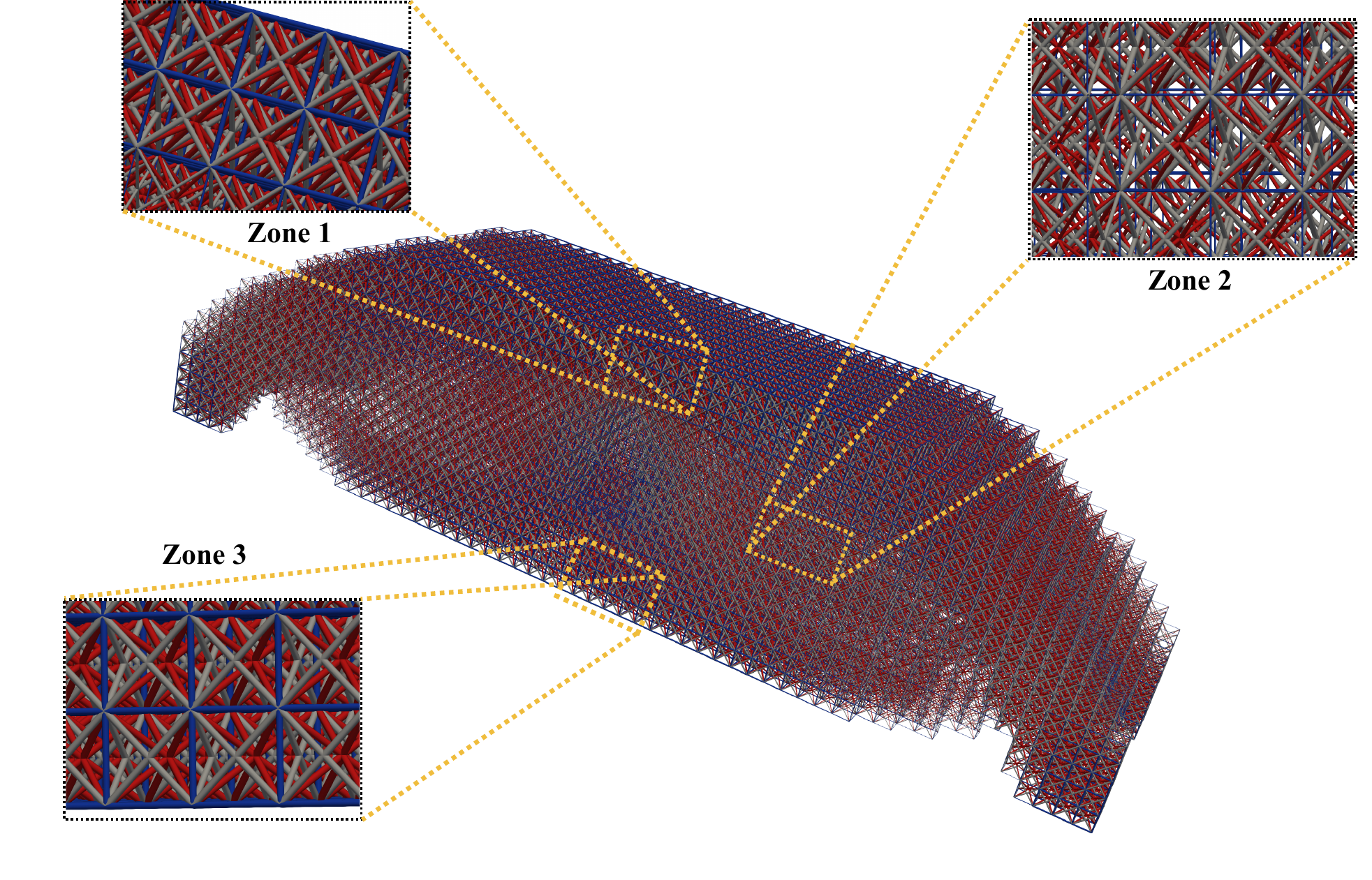}
    \caption{Physical superimposed-lattice strut geometry of the optimized structure. The struts are colored by their respective lattice family: BCC (red), FCC (gray), simple cubic (blue). For better visibility, the diameters are shown at one-fifth of true scale. Three representative regions are highlighted.}
    \label{fig:optimized_structure_zones}
\end{figure*}

To complement the smooth field representation in Fig.~\ref{fig:optimized_fields}, Fig.~\ref{fig:optimized_structure_zones} shows the actual physical superimposed-lattice strut geometry generated from the optimized $a_1$, $a_2$, $a_3$ fields, with strut diameters set from the converged aspect-ratio value at each macroscale element via Eq.~\eqref{eq:aspect_ratio_definition}, and struts colored by lattice family. For visual clarity, strut diameters are shown at one-fifth of their true scale, with relative diameters preserved exactly. The three highlighted regions illustrate how the locally dominant lattice family varies across the structure. In Zones~1 and~2, the BCC (red) and FCC (gray) components have the largest local strut diameters, while the simple cubic component (blue) is comparatively thin. In Zone~3, however, all lattice families take comparable strut diameters. This shows that different lattice families become locally dominant in different regions of the optimized structure rather than contributing equally everywhere.

% ------------------------------------------------------------
\subsection{Discussion}
\label{subsec:discussion}
% ------------------------------------------------------------

The results show that the proposed framework can generate graded microstructural fields while avoiding repeated online homogenization during optimization. The stiffness surrogate provides homogenized stiffness predictions without running new unit-cell simulations, while the density surrogate enables a material constraint based on the actual relative density of the superimposed lattice.

Compared with a standard SIMP-only result, the present framework provides additional microstructural information. The final output is not only a solid--void topology but also a set of spatially varying lattice parameters and corresponding relative density values. These fields can be interpreted as a graded lattice design over the optimized structural domain. The use of a regular superimposed lattice also makes the result easier to interpret. Since $a_1$, $a_2$, and $a_3$ correspond to BCC, FCC, and simple cubic components, respectively, the optimized fields indicate how each lattice family contributes locally to the structural response. This is useful from a design point of view, especially when compared with stochastic or less interpretable microstructure descriptors.

The accuracy of the framework is ultimately limited by that of the surrogate models. The stiffness surrogate reproduces the homogenized Cholesky components with an aggregate relative $L_2$ error of $9.2\,\%$, rising to $13.0\,\%$ for the normal-stiffness factor $G_{33}$. This approximation error enters the assembled stiffness matrix at every design iteration and therefore propagates into the optimized lattice parameter and density fields. The optimized fields should thus be regarded as approximate. A quantitative assessment of this error propagation, for example by comparing the surrogate-based response against direct homogenization of the converged design, is left for future work.

Beyond the coarse manufacturability screen already imposed by bounding $a_i$ to $[0.1, 0.4]$, the current formulation does not enforce further process- or geometry-specific manufacturing constraints. The framework is dimensionless and does not fix an absolute physical scale for the unit cell, so the true manufacturability of the resulting absolute strut diameters remains scale-dependent. However, strut orientation relative to a build direction (overhang angles), the geometric continuity of struts between neighboring graded unit cells, or local buckling of individual struts under load are currently not considered. These aspects are important for future work if the optimized lattice design is to be fabricated using additive manufacturing. A fully concurrent macro--micro formulation would also allow feedback from the microstructure to the macroscale topology and may further improve the final design.

% ============================================================
\section{Conclusion and outlook}
\label{sec:conclusions}
% ============================================================
This work presented a machine-learning-assisted multiscale topology optimization framework for functionally graded superimposed lattice structures. The proposed microscale design space is based on a regular unit cell formed by superimposing BCC, FCC, and simple cubic lattice components, each controlled by an independent geometric parameter. This provides an interpretable design space for spatially tuning the effective stiffness and relative density of lattice-based structures.

Offline computational homogenization was used to generate stiffness data for the proposed lattice family. A Cholesky-constrained neural network surrogate was trained to predict the homogenized stiffness tensor in a physically admissible form, while a separate neural network was trained to predict the relative density using Monte Carlo-based density estimates. These two surrogate models allow stiffness and density evaluations during optimization without repeated online unit-cell homogenization. The surrogate models were integrated into a two-stage optimization strategy. First, SIMP-based topology optimization was used to obtain the macroscale material layout. The resulting solid region was then used for microscale lattice optimization, where the local lattice parameters were updated using the method of moving asymptotes. The framework was demonstrated on a three-dimensional MBB beam benchmark. The computational results show that the proposed method can generate spatially varying distributions of the lattice parameters $a_1$, $a_2$, and $a_3$, together with the corresponding relative density field. The optimized fields are physically interpretable and consistent with the expected load-carrying behavior of the MBB beam. The convergence history also indicates stable microscale optimization while satisfying the prescribed material constraint.

Future work should address the main limitations of our framework discussed in Section~\ref{subsec:discussion} by developing a more strongly coupled macro--micro formulation, incorporating process- and geometry-specific manufacturing constraints, and validating the optimized graded lattices through detailed dehomogenized simulations and experiments. The proposed framework can also be extended to additional unit-cell families and to nonlinear, dynamic, or multiphysics structural responses.

% ============================================================
\section*{CRediT authorship contribution statement}
% ============================================================

\textbf{Prashant Kumar Gupta:} Conceptualization, Methodology, Software, Data curation, Formal analysis, Investigation, Visualization, Writing -- original draft. \textbf{Jonathan Stollberg:} Supervision, Conceptualization, Methodology, Software, Visualization, Writing -- review and editing. \textbf{Dominik Schillinger:} Supervision, Conceptualization, Writing -- review and editing. \textbf{Mohammad Ashraf Iqbal:} Supervision, Conceptualization, Writing -- review and editing.

% ============================================================
\section*{Data availability}
% ============================================================

The data and codes used in this study are available from the corresponding author upon reasonable request.

% ============================================================
\section*{Declaration of competing interest}
% ============================================================

The authors declare that they have no known competing financial interests or personal relationships that could have appeared to influence the work reported in this paper.

% ============================================================
\section*{Acknowledgements}
% ============================================================

Prashant Kumar Gupta gratefully acknowledges the support and guidance received from the Department of Civil Engineering, Indian Institute of Technology Roorkee, and the Department of Civil and Environmental Engineering, Technical University of Darmstadt. He also acknowledges the support received during the research stay at Technical University of Darmstadt.

% ============================================================
\section*{Declaration of generative AI and AI-assisted technologies in the 
manuscript preparation process}
% ============================================================

During the preparation of this work, the authors used Claude Opus 4.8 in order to improve the grammar, language, and readability of the manuscript. After using this tool, the authors reviewed and edited the content as needed and take full responsibility for the content of the published article.

% ============================================================
\appendix
\section{Symmetry relations for the Cholesky factor}
\label{app:symmetry_relations}
% ============================================================

This appendix gives the symmetry relations used to recover the three dependent entries $G_{66}$, $G_{32}$, and $G_{31}$ of the lower triangular Cholesky factor $\mathbf{G}$ of Eq.~\eqref{eq:g_matrix_structure} from the six independent entries predicted by the stiffness surrogate from Section~\ref{subsec:stiffness_surrogate}. Under the tetragonal material symmetry of the superimposed lattice family considered here, the dependent entries follow from
\begin{align}
    G_{66} &= G_{44} \,, \label{eq:g66_symmetry}\\
    G_{32} &= \dfrac{ \dfrac{G_{22}G_{11}}{G_{21}} + \sqrt{ \dfrac{G_{22}^2 G_{11}^2}{G_{21}^2} - G_{33}^2\!\left(1+\dfrac{G_{22}^2}{G_{21}^2}\right) }}{1+\dfrac{G_{22}^2}{G_{21}^2}} \,, \label{eq:g32_symmetry}\\
    G_{31} &= \frac{G_{11}G_{21} - G_{22}G_{32}}{G_{21}} \,. \label{eq:g31_symmetry}
\end{align}
To ensure that the diagonal entries are positive and that the expression inside the square root of Eq.~\eqref{eq:g32_symmetry} is non-negative, the following constraints are imposed:
\begin{align}
    G_{11}^2 > G_{33}^2 + \frac{G_{33}^2 G_{21}^2}{G_{22}^2}\,, \qquad G_{11}, G_{22}, G_{33}, G_{44}, G_{55} > 0 \,.
    \label{eq:positivity_constraints}
\end{align}

\bibliographystyle{elsarticle-num}
\bibliography{references}

\end{document}